\documentclass[aps,prl,twocolumn]{revtex4}
\usepackage{epsf}

\begin{document}

\title {
Upper-critical dimension in a quantum impurity model: \\
Critical theory of the asymmetric pseudogap Kondo problem
}

\author{Matthias Vojta and Lars Fritz}
\affiliation{\mbox{Institut f\"ur Theorie der Kondensierten Materie,
Universit\"at Karlsruhe, 76128 Karlsruhe, Germany}}
\date{March 25, 2004} 

\begin{abstract}
Impurity moments coupled to fermions
with a pseudogap density of states display a quantum phase transition
between a screened and a free moment phase upon variation of the Kondo
coupling.
We describe the universal theory of this transition
for the experimentally relevant case of particle-hole asymmetry.
The theory takes the form of a crossing between effective singlet and
doublet levels, interacting with low-energy fermions.
Depending on the pseudogap exponent, this interaction is either relevant or
irrelevant under renormalization group transformations, establishing
the existence of an upper-critical ``dimension'' in this impurity
problem.
Using perturbative renormalization group techniques
we compute various critical properties and compare with numerical results.
\end{abstract}

\maketitle


Non-trivial fixed points and associated phase transitions in
quantum impurity problems have attracted considerable interest
in recent years,
with applications for impurities in correlated bulk systems, in
transport through nanostructures, and for
strongly correlated lattice models in the framework of dynamical
mean-field theory.
Many of those impurity phase transitions occur in variations
of the well-known Kondo model \cite{hewson} which describes
the screening of localized magnetic moments by metallic conduction
electrons.
A paradigmatic example of an intermediate-coupling impurity fixed point
can be found in the two-channel Kondo effect.

Non-metallic hosts, where the fermionic bath density of states (DOS)
vanishes at the Fermi level, offer a different route to unconventional
impurity physics.
Of particular interest is the Kondo effect in so-called pseudogap
systems \cite{withoff,cassa,tolya2,bulla,GBI,insi,LMA}, described by the Hamiltonian
\begin{equation}
   H = \sum_{k} \varepsilon_k c^\dagger_{k\sigma}
        c_{k\sigma} + J_K {\bf S} \cdot {\bf s}(0)
\label{pgk}
\end{equation}
where ${\bf s}(0)$ is the conduction electron spin at the impurity site,
${\bf S}$ is the spin-$\frac{1}{2}$ impurity, and the rest of the notation
is standard.
Most importantly, the fermionic density of states follows a power law at low energies,
$\rho(\omega)\!=\!\sum_{\bf k} \delta(\omega-\epsilon_{\bf k}) \propto N_0 |\omega|^r$
($r>0$).
Such a behavior arises in semimetals, in certain zero-gap semiconductors,
and in systems with long-range order where
the order parameter has nodes at the Fermi surface, e.g.,
$p$- and $d$-wave superconductors ($r=2$ and 1).
Indeed, in $d$-wave high-$T_c$ superconductors
non-trivial Kondo-like behavior has been observed associated with the magnetic
moments induced by Zn impurities~\cite{bobroff}.
Note that the limit $r\to\infty$ corresponds to a system with a hard gap.

The pseudogap Kondo problem has attracted substantial attention
during the last decade.
A number of studies \cite{withoff,cassa,tolya2}
employed a slave-boson large-$N$ technique; 
significant progress and insight came from
numerical renormalization group (NRG) calculations \cite{bulla,GBI,insi}
and the local moment approach~\cite{LMA}.
It was found that a zero-temperature phase transition occurs at a critical
Kondo coupling, $J_c$, below which the impurity spin is unscreened
even at lowest temperatures.
Also, the behavior depends sensitively on the presence or absence of
particle-hole (p-h) asymmetry, which can arise, e.g., from a band asymmetry at high
energies or a potential scattering term at the impurity site.
A comprehensive discussion of possible fixed points
and their thermodynamic properties has been given in Ref. \onlinecite{GBI}
based on the NRG approach.

So far, analytical knowledge about the critical properties of the pseudogap Kondo
transition is limited.
Previous works employed a weak-coupling renormalization
group (RG) method, based on an expansion in the dimensionless Kondo
coupling $j \!=\! N_0 J_K$.
It was found that an unstable RG fixed point exists at $j\!=\!r$,
corresponding to a continuous phase transition between the free and screened
moment phases \cite{withoff}.
Thus, the perturbative computation of critical properties within this approach
is restricted to small $r$.
Interestingly, the NRG studies \cite{GBI} showed that the
fixed-point structure changes at $r=r^\ast\approx 0.375$
and also at $r\!=\!\frac{1}{2}$, rendering the relevant case of $r\!=\!1$
inaccessible from weak coupling.
Numerical calculations \cite{GBI,insi} also indicated that the critical
fluctuations in the p-h asymmetric case change their character at $r\!=\!1$:
whereas for $r\!<\!1$ the exponents take non-trivial $r$-dependent values and obey
hyperscaling, exponents are trivial for $r\!>\!1$ and hyperscaling is violated.
These findings suggest to identify $r\!=\!1$ as upper-critical ``dimension''
of the problem (whereas $r\!=\!0$ plays the role of the lower-critical
``dimension'').
However, the relevant local degrees of freedom were not known so far.


This letter will present the low-energy theory for the quantum transition
in the p-h asymmetric pseudogap Kondo model, suitable for calculating
critical properties close to and above $r\!=\!1$.
The ingredients are a crossing of many-particle singlet and doublet
levels, and their interaction with the low-energy conduction electrons.
Using RG methods, we show that this
interaction is relevant for $r\!<\!1$ and irrelevant for $r\!>\!1$.
For $r\!<\!1$, we calculate the RG flow of the coupling, and
compute universal critical properties by renormalized perturbation
theory, in a systematic expansion in $(1\!-\!r)$.
For $r\!>\!1$, bare perturbation theory is sufficient, and non-universal
terms appear, as expected for a theory above its upper-critical dimension.


{\it Low-energy field theory.}
It is useful to consider the Kondo problem with a hard-gap
DOS \cite{hardgap}:
in the presence of p-h asymmetry it shows a first-order quantum transition, i.e.,
a level crossing, between a Kondo-screened singlet and a spin-$\frac{1}{2}$ doublet
state.
As we can understand the pseudogap DOS of (\ref{pgk}) as consisting
of an asymmetric high-energy part and a (asymptotically) symmetric
low-energy part, we can obtain an effective theory for (\ref{pgk})
by coupling the above mentioned three (many-body) impurity states,
obtained by integrating out high-energy degrees of freedom from (\ref{pgk}),
to the remaining low-energy part of the conduction electron spectrum.
To represent the three impurity states we introduce
auxiliary operators for pseudoparticles
$b_s$ (for the singlet), $f_\sigma$ with $\sigma=\uparrow,\downarrow$
(for the doublet), with the constraint
$b_s^\dagger b_s + f_\sigma^\dagger f_\sigma = \hat{Q} = 1$.
The form of the coupling to the conduction electrons is dictated by
SU(2) symmetry.

The effective theory to be discussed in the remainder of the paper takes
the form:
\begin{eqnarray}
\label{th}
H &=& (s_0 + \lambda)f_\sigma^\dagger f_\sigma + \lambda b_s^\dagger b_s \\
  &+& g_0 \left(f_\sigma^\dagger b_s c_\sigma(0) + {\rm h.c.}\right)
  \,+ \, \int_{-\Lambda}^{\Lambda} {\rm d}k\,|k|^r \,
  k c_{k\sigma}^\dagger c_{k\sigma}
\nonumber
\end{eqnarray}
where we have represented the bath by linearly dispersing chiral fermions $c_{k\sigma}$,
and $c_\sigma(0) = \int {\rm d}k|k|^r c_{k\sigma}$
is the conduction electron operator at the impurity site.
The spectral density of the $c_\sigma(0)$ fermions
follows the power law $|\omega|^r$ below the
ultraviolet (UV) cutoff $\Lambda$.

The parameter $s_0$ drives the system through the phase
transition at $s_{0c}$.
For $s_0<s_{0c}$ the ground state is a doublet corresponding to
a unscreened local moment (LM), whereas $s_0>s_{0c}$ leads
to a screened singlet state (ASC, see below).
For coupling $g_0=0$ the transition is simply a level crossing,
and $s_{0c}=0$.
As the original Kondo coupling $J_K$ in (\ref{pgk}) determines the
relative positions of the singlet and doublet levels, we
have $(s_0 - s_{0c}) \propto (J_K - J_c)$ near the transition.

The constraint for the Hilbert space of the pseudoparticles, $\hat{Q}=1$,
will be implemented using the chemical potential $\lambda\to\infty$,
such that observables $\langle\hat{\cal O}\rangle$ have to be calculated
according to
$\langle\hat{\cal O}\rangle =
\lim_{\lambda\to\infty} \langle\hat{Q}\hat{\cal O}\rangle_\lambda / \langle\hat{Q}\rangle_\lambda$
\cite{lambda}.

A crucial ingredient is the p-h asymmetry of the original model (\ref{pgk}).
It is clear that upon integrating out the high-energy part of the bath
{\em two} many-body singlet states arise: in the strong-coupling limit they
can be understood as either an electron or a hole bound to the impurity
spin.
Due to the p-h asymmetry of the underlying model these two singlet states will
have very different energies, such that we can discard the high-energy state
in the low-energy theory (\ref{th})
[without loss of generality we have kept the ``electron'' state].
The two singlet states are related by a global p-h transformation,
and they constitute the ground states at the two asymmetric strong-coupling fixed
points (ASC) \cite{GBI} of the pseudogap Kondo model.

Interestingly, the theory (\ref{th}) is formally identical to an infinite-$U$
Anderson impurity model, and we will comment on this connection below.


{\it Scaling analysis and renormalization group.}
Tree level scaling analysis shows that ${\rm dim}[g_0] = \frac{1-r}{2} \equiv {\bar r}$.
Therefore, $r\!=\!1$ plays the role of an upper-critical ``dimension'' where
$g_0$ is marginal.
The case $r\!<\!1$ requires a RG treatment in analogy to the standard
$\epsilon$ expansion of the $\phi^4$ theory.
We shall employ here the familiar momentum shell method
with cutoff $\Lambda$.
For convenience, we introduce masses in each of the $f_\sigma$ and $b_s$
propagators, $m_{0\sigma}$ and $m_{0s}$, keeping in mind that only
their difference $s_0 \equiv m_{0\sigma}-m_{0s}$ has physical significance.
Dimensionless quantities are defined as
$g_0 = \Lambda^{\bar r} g$,
$m_{0\sigma} = \Lambda m_\sigma$, $m_{0s} = \Lambda m_s$.
We set up RG equations for $g$, $m_\sigma$, and $m_s$,
where the fixed-point value of the non-linear coupling $g^2$ will be
of order $\bar{r}$.
Subsequently, we can compute critical properties in a double expansion
in $g$ and $\bar{r}$.

\begin{figure}[t]
\epsfxsize=3.4in
\centerline{\epsffile{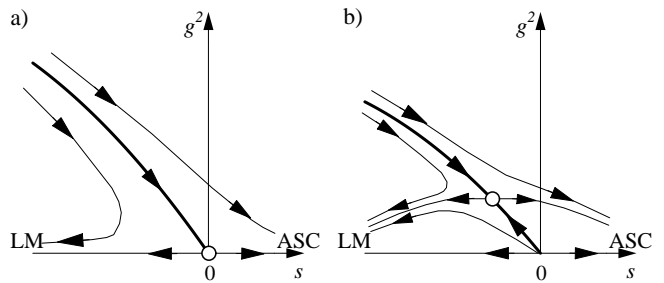}}
\caption{
Schematic RG flow diagram for the effective theory (\protect\ref{th}).
The horizontal axis denotes the energy difference between doublet and
singlet impurity levels, $s$, the vertical axis is the fermionic coupling $g$.
(In the language of the infinite-$U$ Anderson model
the $f$-electron energy is $\epsilon_f\equiv s$, and the hybridization $V\equiv g$.)
The thick lines correspond to continuous boundary phase transitions;
the open circles are the critical fixed points.
a) $r\!>\!1$: $g$ is irrelevant, and the transition is a level crossing with perturbative corrections.
b) $r\!<\!1$: $g$ is relevant, and the transition is controlled by an interacting
fixed point at ${g^\ast}^2 = (1-r) /3 + {\cal O}(1-r)^2$ (\protect\ref{fp}).
}
\label{fig:flow}
\end{figure}

To one-loop order we obtain the following RG beta functions:
\begin{eqnarray}
\beta(g) &=& - {\bar r} g + \frac{3}{2} g^3
\,, \nonumber \\
\beta(s) &=& - s + 3 g^2 s - g^2
\label{beta}
\end{eqnarray}
where $s = m_\sigma - m_s$.
For $r<1$ the trivial fixed point $g^\ast=s^\ast=0$ is unstable,
and the critical properties are instead controlled by an interacting
fixed point at
\begin{equation}
{g^\ast}^2 = \frac{2 \bar r}{3} \,,~~
s^\ast = - \frac{2 \bar r}{3} \,.
\label{fp}
\end{equation}
At this fixed point, we find anomalous field dimensions
$\eta_s = 2 {g^\ast}^2$, $\eta_\sigma = {g^\ast}^2$, and
a correlation length exponent
\begin{equation}
\nu z = (1 - 3 {g^\ast}^2)^{-1} = 1/r + {\cal O}(\bar{r}^2)\,,
\label{nuz}
\end{equation}
describing the vanishing of the characteristic crossover
temperature in the vicinity of the critical point,
$T^\ast \propto |s_0-s_{0c}|^{\nu z}$ \cite{book}.

The resulting RG flow diagram is shown in Fig.~\ref{fig:flow}.
(No qualitative changes occur to two-loop order.)
We have also indicated the stable fixed points
reached for $s\to\pm\infty$:
Large positive $s$ implies a non-degenerate singlet ground state of the impurity (ASC),
where the interaction $g_0$ is clearly irrelevant.
Large negative $s$ describes a doublet impurity state (LM);
for small $g_0$ one can perform the usual Schrieffer-Wolff transformation
to project out the singlet state.
This results in a pseudogap Kondo model of the form (\ref{pgk}), with
$J_K \propto g_0^2/|s_0|$ --
here small $J_K$ is irrelevant for $r>0$ \cite{withoff,MVMK}.

The critical fixed point (Fig. \ref{fig:flow}b) shifts to larger values of
${g^\ast}^2$, $|s^\ast|$ with decreasing $r$.
In the light of the numerical results of Ref.~\onlinecite{GBI}
we expect that Eq. (\ref{th}) together with an expansion in $g$, $\bar{r}$
yields the correct description of the critical properties for
$0.375\approx r^\ast < r < 1$.
Using NRG we have numerically confirmed this expectation, i.e., the properties of
the critical fixed point of the model (\ref{th}) vary continuously as function
of $r$ for $r^\ast < r < 1$.
Considering that Eq. (\ref{th}) represents an infinite-$U$ Anderson model,
we also conclude that in this $r$ range the phase transition of the pseudogap
Anderson model \cite{GBI} is in the same universality class as the one of
the pseudogap Kondo model.

\begin{figure}[!t]
\epsfxsize=2.9in
\centerline{\epsffile{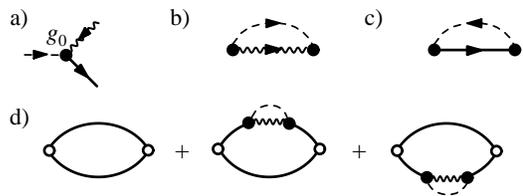}}
\caption{
Feynman diagrams occurring in the perturbation theory for (\protect\ref{th}).
Full/wiggly/dashed lines denote $f_\sigma$/$b_s$/$c_\sigma$ propagators.
a) Bare interaction vertex $g_0$.
b) $f_\sigma$ self-energy.
c) $b_s$ self-energy.
d) Diagrams entering the local susceptibility, needed to obtain the anomalous
exponent $\eta_\chi$ (\protect\ref{eta_chi}).
}
\label{fig:dgr}
\vspace*{-10pt}
\end{figure}

We proceed by calculating a few observables in the vicinity of the
critical point for $r<1$ using standard renormalized perturbation theory;
diagrams are displayed in Fig.~\ref{fig:dgr}.
Power counting shows that the integrals appearing in universal quantities
will display no UV singularities, thus the cutoff $\Lambda$ can be sent to infinity.
The structure of the critical theory implies hyperscaling properties,
including $\omega/T$ scaling in dynamical quantities \cite{book}.
The local dynamic susceptibility follows a scaling form
\begin{equation}
\chi''_{\rm loc}(\omega,T)
= \frac {{\cal B}_1} {\omega^{1-\eta_{\chi}}} \, \Phi_1 \!\left(\frac{\omega}{T}, \frac{T^{1/(\nu z)}}{s_0-s_{0c}}\right)
\,,
\label{imchi}
\end{equation}
where $\Phi_{1}$ is a universal crossover function (for fixed $r$), and
${\cal B}_{1}$ is a non-universal prefactor.
A similar scaling forms holds for the local static susceptibility $\chi_{\rm loc}(T)$,
and at the critical point $\chi_{\rm loc}(T) \propto T^{-1+\eta_\chi}$.
The anomalous exponent $\eta_\chi$ can be calculated in an expansion in $\bar{r}$,
\begin{equation}
\eta_\chi =  \frac{4}{3}\bar r + {\cal O}(\bar{r}^2) \,,
\label{eta_chi}
\end{equation}
which fits the numerical findings, e.g., $1-\eta_\chi = 0.928\pm0.002$
for $r=0.9$ \cite{insi}.
Using hyperscaling, the results (\protect\ref{nuz}) and (\ref{eta_chi})
are sufficient to determine all critical exponents associated with
a local magnetic field \cite{insi}, in an expansion in $\bar{r}$.
In particular, the $T\to 0$ local susceptibility away from criticality
obeys:
\begin{eqnarray}
\chi_{\rm loc}(s_0\!>\!s_{0c})  &\propto& (s_{0}-s_{0c})^{-\gamma} \,,~
\gamma = \nu z \,(1-\eta_\chi) \,,\nonumber\\
T\chi_{\rm loc}(s_0\!<\!s_{0c}) &\propto& (s_{0c}-s_{0})^{\gamma'} \,,~
\gamma' = \nu z \, \eta_\chi \,.
\end{eqnarray}
As proposed earlier \cite{GBI,insi}, $\lim_{T\to 0}T\chi_{\rm loc}$ can serve as an order
parameter, as it vanishes continuously as $J_K\to J_c^-$, and
is zero for $J_K>J_c$.

A scaling form similar to (\ref{imchi}) holds for the conduction electron $T$ matrix,
describing the scattering of the $c$ electrons off the impurity.
At the critical point the $T$ matrix will obey
$T(\omega) \propto \omega^{-1+\eta'_T}$.
As in Ref.~\onlinecite{MVMK} we are able to determine an {\em exact} result for
the anomalous exponent $\eta'_T$ \cite{texp}, valid to all orders in perturbation theory.
The argument is based on the diagrammatic structure of the $T$ matrix:
the relevant diagrams can be completely constructed from full $g$
vertices and full $f/b$ propagators \cite{MVMK}.
We obtain the exact result:
\begin{equation}
\eta'_T= 2\bar{r} ~~~\Rightarrow~~~
T(\omega) \propto \omega^{-r}.
\label{exact}
\end{equation}
Note that the similar result of Ref.~\onlinecite{MVMK} was derived
for the p-h symmetric critical fixed point of the model (\ref{pgk}),
and is valid for $0<r<\frac{1}{2}$.
In contrast, Eq. (\ref{exact}) applies to the p-h asymmetric
fixed point (\ref{fp}), and is valid for $r^\ast<r<1$.
NRG calculations have found precisely this critical divergence,
${\rm Im}\,T(\omega)\propto |\omega|^{-r}$, for both the symmetric
and asymmetric critical points \cite{bulla,MVRB}.
Notably, the $T$ matrix can be directly observed in experiments,
due to recent advances in low-temperature scanning tunneling microscopy,
as has been demonstrated, e.g., with high-temperature
superconductors \cite{seamus,tolya,MVRB}.

Lastly, we mention that it is also possible to compute the impurity
residual entropy at the critical point, by expanding the free energy
in $g$. The result is
\begin{equation}
S_{\rm imp} (T\!=\!0) = \ln 3 - \frac{16 \ln 2}{9}\,{\bar r} + {\cal O}(\bar{r}^2)\,,
\end{equation}
again in agreement with numerics, Fig. 14 of Ref.~\onlinecite{GBI}.


{\it Perturbation theory for $r>1$.}
For bath exponents $r\!>\!1$ the coupling $g_0$ in the theory (\ref{th})
is irrelevant in the RG sense, and therefore observables can be obtained
by straightforward perturbation theory.
Note that the cutoff $\Lambda$ has to be kept explicitly, as integrals
will be UV divergent.
This also implies that hyperscaling is violated, and no $\omega/T$ scaling
in dynamics occurs, as usual for a theory above the upper-critical dimension.

We shall demonstrate the calculation of the local impurity
susceptibility $\chi_{\rm loc}$.
To lowest non-trivial order, $\chi_{\rm loc}$ is given by the convolution
of two $f_\sigma$ propagators, calculated with the self-energy to second order in $g_0$.
Note that no vertex corrections occur to this order due to the
structure of the interaction.
The self-energy, Fig. \ref{fig:dgr}b, is
\begin{eqnarray}
\Sigma_{f_\sigma}(i\omega_n) = \frac{g_0^2 T}{(2\pi)^{1+r}} \sum_{i\omega'_n}
\int \frac{{\rm d}k\,|k|^r}{i\omega'_n\!-\!k} \,
\frac{1}{i\omega_n \!-\! i\omega'_n \!-\!\lambda\!-\!m_{0s}} \,.
\nonumber
\end{eqnarray}
The real part of this self-energy is $\propto\Lambda$ and will renormalize the mass
$m_{0\sigma}$, this shift can be absorbed by defining
$
\bar{m}_\sigma = m_{0\sigma} + {\rm Re} \Sigma_{f_\sigma}(\lambda+\bar{m}_\sigma)
$,
similarly for $\bar{m}_s$. Subsequently, all quantities are expressed in terms of
$\bar{m}_\sigma$, $\bar{m}_s$, and
$\bar{s} \equiv s_0-s_{0c} = \bar{m}_\sigma -\bar{m}_s$ measures the distance to the
transition point.

At the transition, $\bar{s}\!=\!0$, the self-energy shows threshold behavior at $T\!=\!0$,
$-{\rm Im}\,\Sigma_{f_\sigma}(\bar{\omega}+i\eta)/\pi \propto g_0^2 \bar{\omega}^r \Theta(\bar{\omega})$
with $\bar{\omega} = \omega-\lambda-\bar{m}_\sigma$.
The low-energy behavior of the $f_\sigma$ propagator follows as
\begin{eqnarray}
-{\rm Im} \, G_{f_\sigma}(\bar{\omega}+i\eta)/\pi =
(1-A) \delta(\bar{\omega}) + B |\bar{\omega}|^{r-2} \Theta(\bar{\omega})
\nonumber
\end{eqnarray}
with $A,B \propto g_0^2$.
The $b_s$ propagator has a similar form.
When calculating $\chi_{\rm loc}$ the $T\to 0$ limit has to be taken with
care, as the exponentially small tail in ${\rm Im} G_{f_\sigma}(\bar{\omega})$ at
$\bar{\omega}<0$ contributes to $\chi_{\rm loc}$.
One obtains for the imaginary part $\chi''_{\rm loc}$ for $T\to 0$ and $1<r<2$:
\begin{equation}
\chi''_{\rm loc}(\omega)/\pi = \frac{1-2A}{6} \, \frac{\delta(\omega)\omega}{T} +
\frac{B}{3} \, |\omega|^{r-2} {\rm sgn}(\omega) \,.
\end{equation}
The static local suceptibility at criticality is easily seen to follow
$\chi_{\rm loc}(T) \propto 1/T$, implying $\eta_\chi=0$, in contrast to
the result for $r<1$ (\ref{eta_chi}).
The impurity residual entropy receives no corrections from the coupling
to the conduction electrons, thus $S_{\rm imp} (T\!=\!0) = \ln 3$.

Moving away from criticality, one of the $f_\sigma$, $b_s$ propagators looses
its $\delta(\bar{\omega})$ contribution --
for $\bar{s}>0$ ($J_K>J_c$) this is $G_{f_\sigma}$, indicating that free-moment
behavior is absent in this regime.
Consequently, the zero-temperature static local susceptibility is finite,
but diverges upon approaching the critical point according to
$\chi_{\rm loc} \propto \bar{s}^{r-2}$.
For $\bar{s}<0$, $\chi_{\rm loc} \propto 1/T$.
Thus, the order parameter $T\chi_{\rm loc}$ jumps at the transition point
for $r>1$.

At $r\!=\!1$, relevant for the $d$-wave superconductor,
the quoted power laws remain valid, but are supplemented by logarithmic
corrections \cite{cassa,GBI,MVRB}, as the coupling $g_0$ is marginal in
this case. We have found the leading corrections to be
$\chi_{\rm loc} \propto 1/(\omega \ln^{2/3} |\omega/\Lambda|)$ and
${\rm Im}\,T(\omega) \propto 1/(|\omega| \ln^2 |\omega/\Lambda|)$.


{\it Conclusions.}
We have developed an effective low-energy theory describing
the quantum phase transition in the asymmetric pseudogap Kondo
model -- this was made possible by identifying the local degrees
of freedom relevant for the transition near its upper-critical
``dimension''.
For bath exponents $r\!<\!1$ we have analyzed the interacting
field theory using perturbative RG techniques,
and established the existence of a transition obeying
universal local non-Fermi liquid behavior and
strong hyperscaling properties including $\omega/T$ scaling of
dynamical quantities.
For $r\!>\!1$ we find a level crossing with perturbative
corrections.
Notably, the RG near the upper-critical dimension ($r\!=\!1$) needed to be
formulated in variables completely different from the ones appropriate near
the lower-critical dimension ($r\!=\!0$).
Our results are in excellent agreement with numerical data;
they are applicable to impurity moments in unconventional
superconductors \cite{bobroff,MVRB,tolya} and other pseudogap systems.
Further, the pseudogap Kondo problem permits a sharp transition upon
application of a local magnetic field, this will be detailed elsewhere.

We expect that field theories similar to ours can be constructed for
other impurity transitions, and will also be useful for
the study of lattice models in dynamical mean-field theory and
its extensions, with applications to
local quantum criticality, discussed both in heavy-fermion metals \cite{edmft}
and high-$T_c$ superconductors \cite{pana}.



We thank R. Bulla, K. Ingersent, M. Kir\'{c}an, A. Rosch, S. Sachdev, Q. Si,
and P. W\"olfle for discussions.
This research was supported by the DFG through
the Center for Functional Nano\-structures Karlsruhe.



\begin{thebibliography}{}
\bibitem{texp}
In the weak-coupling approach of Ref. \protect\onlinecite{MVMK} the anomalous exponent of the
$T$ matrix is defined differently.
\end{thebibliography}
\end{document}